\documentclass[twocolumn,preprintnumbers,superscriptaddress,amsmath,amssymb,nofootinbib,aps]{revtex4}

\usepackage{graphicx}
\usepackage{slashed}
\usepackage{bbm}
\usepackage{footmisc}
\usepackage{subfigure}
\usepackage{times,color}

\newcommand{\be}{\begin{eqnarray*}}
\newcommand{\ee}{\end{eqnarray*}}
\newcommand{\gl}[1]{(\ref{#1})}

\newcommand{\bee}{\begin{eqnarray}}
\newcommand{\eee}{\end{eqnarray}}
\newcommand{\beeq}{\begin{equation}}
\newcommand{\eeeq}{\end{equation}}
\newcommand{\gev}{~{\text{GeV}}}
\newcommand{\citeh}[1]{\hbox{\cite{#1}}}

\renewcommand{\vec}{\bf}

\begin{document}

\title{Partially (in)visible Higgs decays at the LHC}

\begin{abstract}
  Both Atlas and CMS have reported a discovery of a Standard
  Model-like Higgs boson $H$ of mass around 125 GeV. Consistency with
  the Standard Model implies the non-observation of non-SM like decay
  modes of the newly discovered particle. Sensitivity to such decay
  modes, especially when they involve partially invisible final states
  is currently beyond scrutiny of the LHC. We systematically study
  such decay channels in the form of
  $H\rightarrow AA\rightarrow \text{jets}+\text{missing energy}$,
  with $A$ a light scalar or scalar, and analyze to what extent these
  exotic branching fractions can be constrained by direct measurements at
  the LHC. While the analysis is challenging, constraints as good as
  ${\text{BR}}\lesssim 10\%$ can be obtained.
\end{abstract}

\preprint{IPPP/12/67, DCPT/12/134}

\author{Christoph Englert} 
\affiliation{Institute for Particle Physics Phenomenology, Department
  of Physics,\\Durham University, DH1 3LE, United Kingdom}
\author{Michael Spannowsky} 
\affiliation{Institute for Particle Physics Phenomenology, Department
  of Physics,\\Durham University, DH1 3LE, United Kingdom}
\author{Chris Wymant} 
\affiliation{Institute for Particle Physics Phenomenology, Department
  of Physics,\\Durham University, DH1 3LE, United Kingdom}

\preprint{}

\maketitle


\section{Introduction}
Recent results obtained at the Large Hadron Collider by the Atlas and
CMS experiments \citeh{atlas,cms} have revealed the existence of a
light Higgs candidate~\citeh{Higgs} with a mass of $\sim 125$~GeV. The
observation of this new particle combines evidence in the
$H\to\gamma\gamma$, $H\to ZZ$ and $H\to W^+W^-$ channels. Given the
yet small collected luminosity, the properties of this newly
discovered resonance are still subject to large statistical
uncertainties~\citeh{fits}. Analyses targeting {\it{e.g.}}  spin and
${\cal{CP}}$~\citeh{cp,cp2} of the new particle and a more precise
extraction of its couplings to known matter will be addressed with a
larger data sample.

The current observations leave open a plurality of phenomenological
possibilities of Higgs
sector-modifications~\citeh{new_ops,Bock:2010nz}. Especially the
extraction of the resonance's couplings is influenced by non-standard
decays~\citeh{evasive,thresh,invis,invis2,invis3} since it is based on
a fit to combinations of various production $p$ and decay modes
$d$. These are functions of the partial and total decay widths and
{\emph{all}} couplings $\{ g_i \}$:
\begin{equation}
  \label{decaywidth}
  \sigma_p \times \text{BR}_d \sim {\Gamma_p\,\Gamma_d\over
    \Gamma_{\text{tot}}}\sim {g_p^2\,g_d^2\bigg/ \bigg(
    \sum_{\text{modes}} g_i^2 \bigg)} \,.
\end{equation}
The total width $\Gamma_{\text{tot}}$ in current fits is typically
approximated by including a freely flowing invisible partial width
\cite{shrock} to the list of decay modes or by imposing the constraint
$\Gamma_{\text{invis}}\sim g_{\text{invis}}^2=0$.

\bigskip

Extracting such an invisible\footnote{Note that from
  Eq.~\gl{decaywidth} ``invisible'' also means fully visible in a
  non-standard search channel.} or partially visible branching ratio
is an experimentally ambitious task. The decay of the Higgs via a
light scalar or pseudo-scalar $A$ can be buried in a large
hadronic background, experimental systematics can limit the
sensitivity to such decays, and non-SM phenomenology can easily be
missed. The signature $H\to AA$ occurs in many extensions of the SM,
{\it{e.g.}}, the next-to-minimal supersymmetric Standard
Model~\citeh{nmssm}, Higgs-portal models~\citeh{Patt:2006fw}, and
whenever an approximate symmetry of the Higgs potential is
explicitly broken by a small term in the potential, giving a light
pseudo-Nambu-Goldstone boson (see~\citeh{Lisanti:2009uy} for a
clear discussion). However the signature can be missed by standard searches,
depending on how $A$ itself decays, and new dedicated strategies need to
be devised. Obviously, an observation of such a novel decay channel
$H\to AA$ would directly imply physics beyond the Standard Model.

Subjet methods (pioneered in Ref.~\citeh{subjethiggs}) have proven
particularly successful in getting a handle on such a modified
phenomenology. In particular, subjet analyses applied to the decay
chains $H\to 2A\to 4X$ have unravelled potential sensitivity to these
non-standard decays if the (pseudo)scalar $A$ is light
${\cal{O}}(10-20\gev)$. A decay of the Higgs to $AA$ is well motivated
on general grounds; one can keep a reasonably open mind with regards
to how the $AA$ subsequently decay. $H\to AA\to 4 g$ is considered
in~\citeh{Falkowski:2010hi}, $H\to AA\to 4\tau$
in~\citeh{Englert:2011iz}, $H\to 2A\rightarrow 2\tau 2\mu$
in~\citeh{Lisanti:2009uy} and $H\to 2A\rightarrow 4c$
in~\citeh{Lewis:2012pf}. Higgs decays into resonances with masses
close to hadronic bound states have been studied in
Refs.~\citeh{thresh,thresh2}. Sensitivity to the signatures discussed
in
Refs.~\citeh{thresh,Lisanti:2009uy,Falkowski:2010hi,Englert:2011iz,Lewis:2012pf,thresh2}
follows from fundamentally distinct QCD and electroweak properties,
highlighting the diverse power that jet substructure-based approaches
offer.

There is one major difference in the analysis of purely hadronic final
states compared to electroweak final states. Quite often the latter
involve a significant amount of missing energy, which is aligned with
the direction of the fat jet that is input to grooming and/or tagging
algorithms~\citeh{boostpros1011}. Normally such an event topology is
avoided to minimize systematic uncertainties. In, {\it{e.g.}},
searches for supersymmetry in the jets+missing energy
channel~\citeh{exisusy} one requires a missing energy
vector $\slashed{E}_T$ well isolated from a number of hard jets, to reduce
systematics. Decays $H\to AA$ with $m_H/m_{A}\gg 2$, on the other
hand, naturally involve non-isolation of $\slashed{E}_T$, which might
even be not too large depending on the decay of $A$. In the phase
space region where we can separate signal from background a
SUSY-inspired search strategy based on a strong $\slashed{E}_T$
isolation becomes impossible.

\begin{figure*}[!t]
  \begin{center}
    \includegraphics[width=0.8\textwidth]{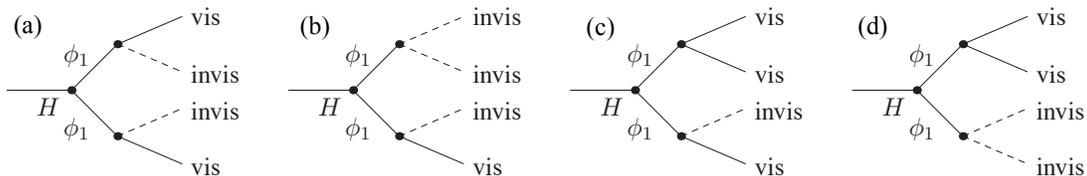}
  \end{center}
  \caption{\label{fig:topref} Higgs decay topologies in the simplified
    models that we study for the purpose of this paper. ``vis'' is a
    placeholder for $u\bar u,d \bar d$ flavor quark pairs that give
    rise to visible hadronic energy.}
\end{figure*}

\bigskip

Strategies to deal with non-SM Higgs decays involving both missing and
hadronic energy \cite{pivis} are, hence, limited. A direct generalization of the
standard invisible Higgs decay search in the weak boson fusion
channel~\citeh{invis} is not possible. Such an analysis relies on a
central jet veto to obtain a sufficiently large signal-over-background
ratio $S/B$ to study the jets' azimuthal angle correlation. Applying a
central jet veto to central decay $H\to {\slashed{E}}_T$+hadrons
removes the signal that we would like to investigate.

Adapted Higgs+monojet searches \cite{evasive,invis3} are challenged by
overwhelmingly large dijet and weak boson+jet backgrounds and trigger
issues, as soon as missing energy from invisible Higgs decays
decreases when turning to a partially visible decay. The latter can at
least be partially cured by focussing on the Higgs'
$p_T$-distribution's tail. This, however, comes at the price of larger
theoretical and experimental uncertainties, which imperatively need to
be included to reach a realistic formulation of the branching ratio
constraints.

This leaves Higgs-strahlung $pp\to HZ$ as the best-motivated channel
to study the situation we have in mind at the LHC. Trigger issues are
avoided by reconstructing the leptonic $Z\to e^+e^-,\mu^+\mu^-$ decay
and no further adjustments to the trigger settings or thresholds are
necessary to perform the measurement(this process is also studied in
the context of invisible Higgs searches~\citeh{Godbole}). Furthermore, the
jet energy scale calibration is performed with $Z$+jet events, whose
distribution is both theoretically and experimentally under good
control~\citeh{calib,missing}. Data-driven methods can be
straightforwardly applied in ``ABCD'' approaches, when {\it{e.g.}}
comparing boosted to the un-boosted $Z$ boson kinematics.

\section{Decay topologies}

We employ a simplified-model based approach~\citeh{simpmo} to
investigate the LHC search potential to the $H\to
{\slashed{E}}_T+$hadrons signature. More precisely we study the Higgs
decay realizations depicted in Fig.~\ref{fig:topref}. These scenarios
(a)-(d) are characterized by different kinematics and different
relative contributions of missing and hadronic energy in the final
state. 

We limit our analysis to the light flavor final states $u,d$;
particles with $\sim 10~\gev$ masses and a significant coupling to $b$
quarks are not only constrained by upsilon
measurements~\citeh{Dermisek:2006py}, but also give rise to plethora
of dedicated phenomenological handles on the final
state~\citeh{thresh2,giac,batell} that we do not wish to exploit to be as
general as possible. Similar arguments hold for the $H\to 4\tau$ decay,
which gives rise to sparse but focused hadronic energy
deposits~\citeh{Englert:2011iz}. Both these cases (and also $H\to 4
g$) are fundamentally different from the topologies of
Fig.~\ref{fig:topref} where ``visible'' refers to light flavor
quarks. The quarks undergo normal showering and hadronization leaving
neither the possibility for flavor tags nor for detecting pronged
decays from counting charged tracks, as done in the $b$- and
$\tau$-flavored decays of $H\to AA$, respectively.

\section{Elements of the Analysis}
\label{sec:analysis}
We implement the decay topologies using
{\sc{FeynRules}}~\citeh{feynrules} and use its interface to
{\sc{Sherpa}}~\citeh{frsherpa,sherpa} to generate events for the mass
choices $m_H=125~\gev$, $m_A=20~\gev$,
$m_{\text{invis}}=10~\gev$. This choice is not special and the details
of eventually extracting the branching ratio is not sensitive to the
particular value $m_A$ unless $m_H\not\gg m_A$. We will comment on the
possibility to extract $m_A$ in Sec.~\ref{sec:bounds}.

We generate background events using {\sc{Sherpa}} and include
$WW$+jet, $WZ$+jet, $ZZ+$jet, and $t\bar t$+jets as the main
backgrounds to our $pp \to (Z\to
\ell^+\ell^-)+{\text{jet}}+{\slashed{E}}_T$ analysis. We normalize our
signal and background event samples to the corresponding higher
order-corrected cross sections \cite{nlo1,nlo2,higgs}. Studying the
impact of a mismeasurement of $Z$+jets events requires the simulation
of a realistic detector environment, and should be addressed by the
experiments. However, we may assume on the basis of
Refs.~\cite{calib,missing} that this background can be brought under
sufficient control and can be subtracted from the eventual
distribution also when the missing energy vector is collinear to the
jet. We include a flat shape uncertainty of the background distribution
to partly account for the jet energy scale uncertainty in the
computation of the expected BR limits in Sec.~\ref{sec:bounds}.

Associated Higgs production with SM-like Higgs decays to $b\bar b$ and
$\tau\tau$ also comprise backgrounds to our $(Z\to
\ell^+\ell^-)+{\slashed{E}_T}+$jet analysis and we include them
consistently throughout.

\begin{figure}[t]
  \includegraphics[width=0.4\textwidth]{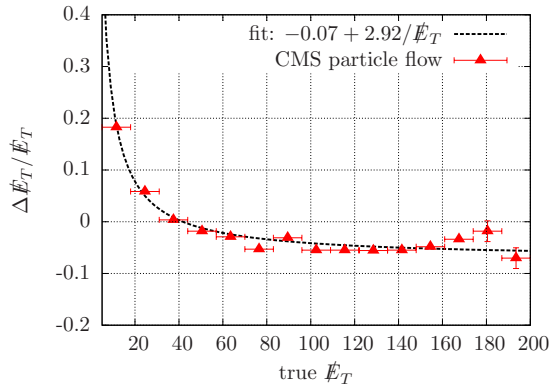}
  \caption{\label{fig:reco} $\Delta{\slashed{E}}_T/{\slashed{E}}_T$ as
    a function of the true $\slashed{E}_T$ that follows from particle
    flow and early LHC data~\citeh{pflow}. We also display the fitted
    function that is employed for our analysis.}
\end{figure}
\begin{figure*}[!t]
  \begin{center}
    \subfigure[][]{
      \includegraphics[width=0.4\textwidth]{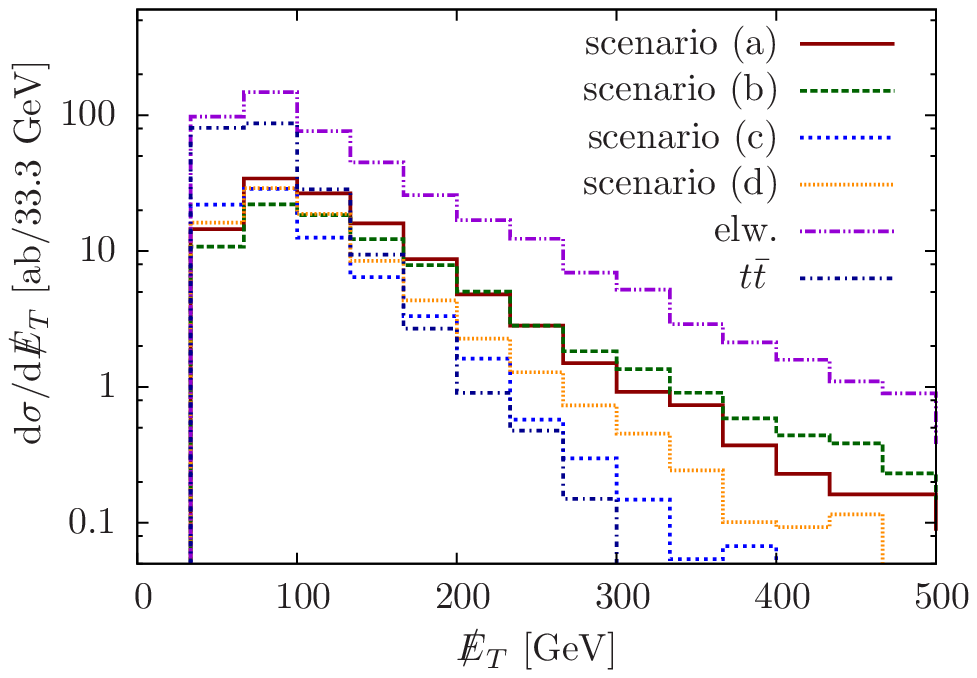}}
    \hspace{1.5cm}
    \subfigure[][]{
    \includegraphics[width=0.4\textwidth]{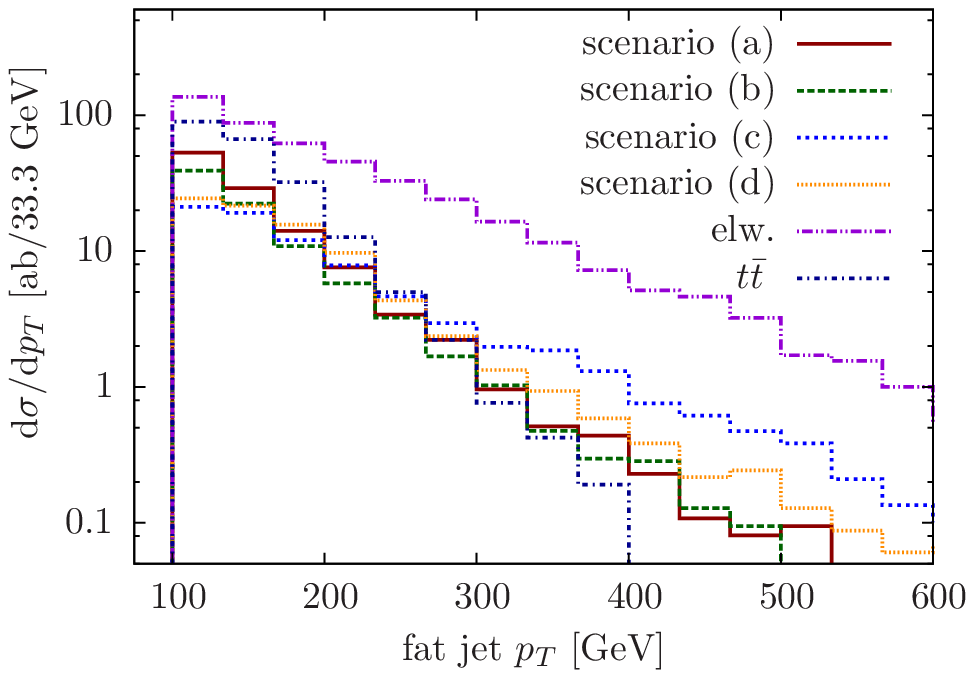}}
  \end{center}
  \caption{\label{fig:backgrounds} Missing transverse energy (a) and
    fat jet transverse momentum (b) of the decay scenarios
    Fig.~\ref{fig:topref}(a)-(d) and the contributing backgrounds
    after all analysis steps have been applied.}
\end{figure*}

\bigskip

We reconstruct the events' visible final states using a hybrid
ECAL+HCAL implementation which granularizes the final state particles
on grids with $\Delta\eta\times \Delta\phi=0.025 \times 0.025~(0.1
\times 0.1)$ ECAL (HCAL) as massless cell entries and feed the
reconstructed objects to a smearing routine which mimics detector
effects as described later on. Doing so, we reconstruct the full
three-momenta from the detector geometry with an invariant mass
$p^2=0$ for each ECAL+HCAL cell, which contains a single or multiple
hits.

In the analysis, we first reconstruct isolated stable leptons, by
requiring the hadronic energy deposit in the vicinity of the lepton
candidate ($\Delta R=\sqrt{\Delta \eta^2 + \Delta\phi^2}\leq 0.1$,
where $\eta$, and $\phi$ are pseudorapidity and azimuthal angle
respectively) to be smaller than 10\% of the lepton candidate's
transverse momentum. We furthermore require exactly two leptons
($p_{T,\ell}>10~\gev,~|\eta_\ell|<2.5$) of identical flavor and
opposite charge that recombine the $Z$ mass within $m_Z\pm 10$~GeV
(note that this also removes higher-pronged $H\to \tau\tau$ events)
and demand $p_{T,Z}=(p_{\ell_1}+p_{\ell_2})_T>130~\gev$.
Subsequently, we cluster Cambridge/Aachen jets with $R=1.5$ using
{\sc{FastJet}}~\citeh{fastjet} and we require at least one such fat
jet with $p_T>100~\gev$. We apply the mass drop and asymmetry criteria
to this jet as explained in Ref.~\citeh{subjethiggs}, we filter and
trim it~\citeh{trim}, and we keep the three hardest
subjets~\citeh{subjethiggs}. Furthermore, the jet needs to pass a
$b$-veto in $|\eta|<2.5$, for which we choose a working
point~\citeh{btagmu} with a flat 70\% tagging efficiency and a flat
10\% fake rate. The $b$ veto removes the pollution from $H,Z\to b\bar
b$ and significantly reduces the $t\bar t$ background.

In the next step we remove the fat jet from the event and re-cluster
the remaining tracks to anti-$k_T$ jets with $R=0.4$ and
$p_T>30~\gev$.  In order to project out the signal region we are
interested in, we require a large missing transverse energy
$E_T>50~\gev$, which is reconstructed from all visible hits.  It
needs to be isolated from the reconstructed $Z$ boson
$R({\slashed{E}_T},Z)\geq 2$, and we veto additional jets around the
$Z$: $\forall j:~R(Z,j)>1.5$ to further reduce $t\bar t$ events.

\begin{figure*}[!t]
  \begin{center}
    \subfigure[][\label{fig:scen1}]{
       \includegraphics[width=0.4\textwidth]{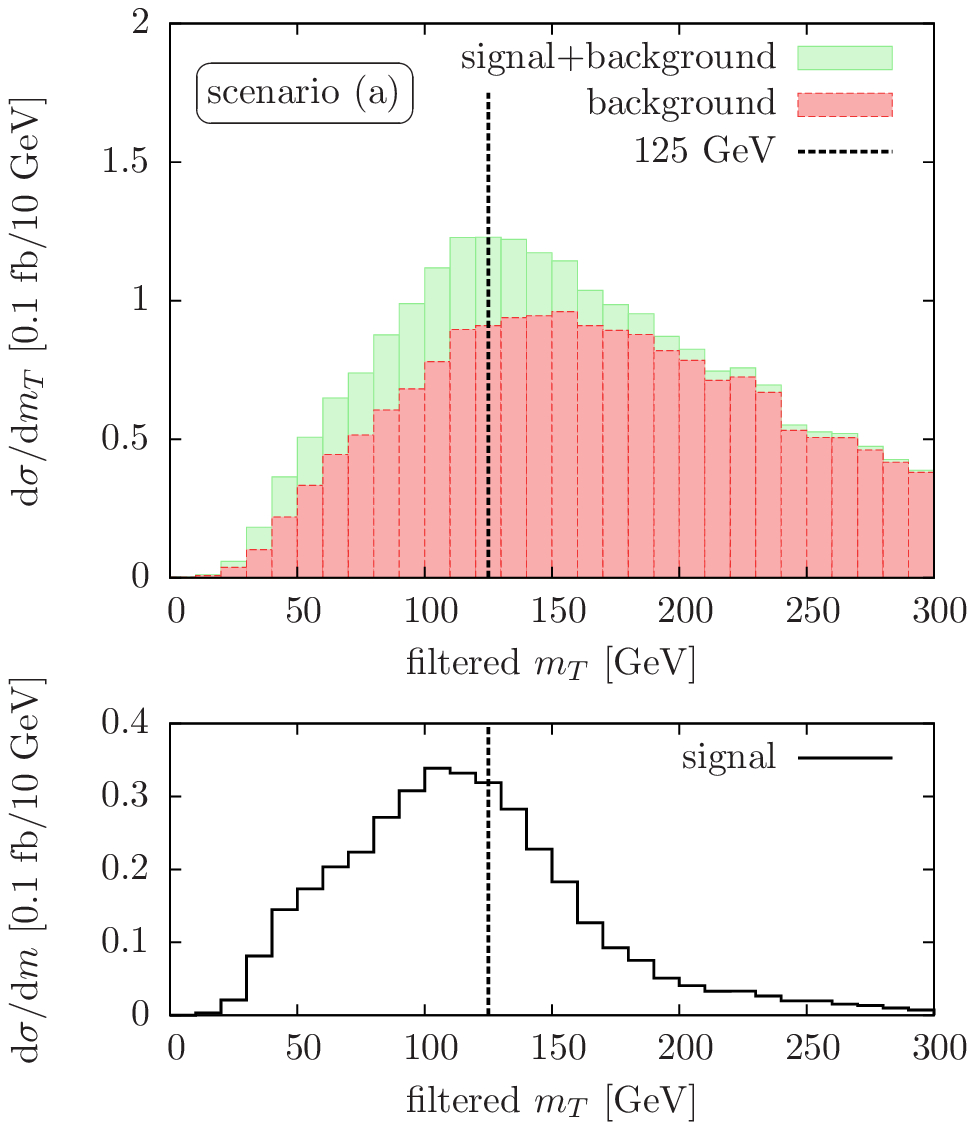}
     }
     \hspace{1.5cm}
    \subfigure[][\label{fig:scen2}]{
      \includegraphics[width=0.4\textwidth]{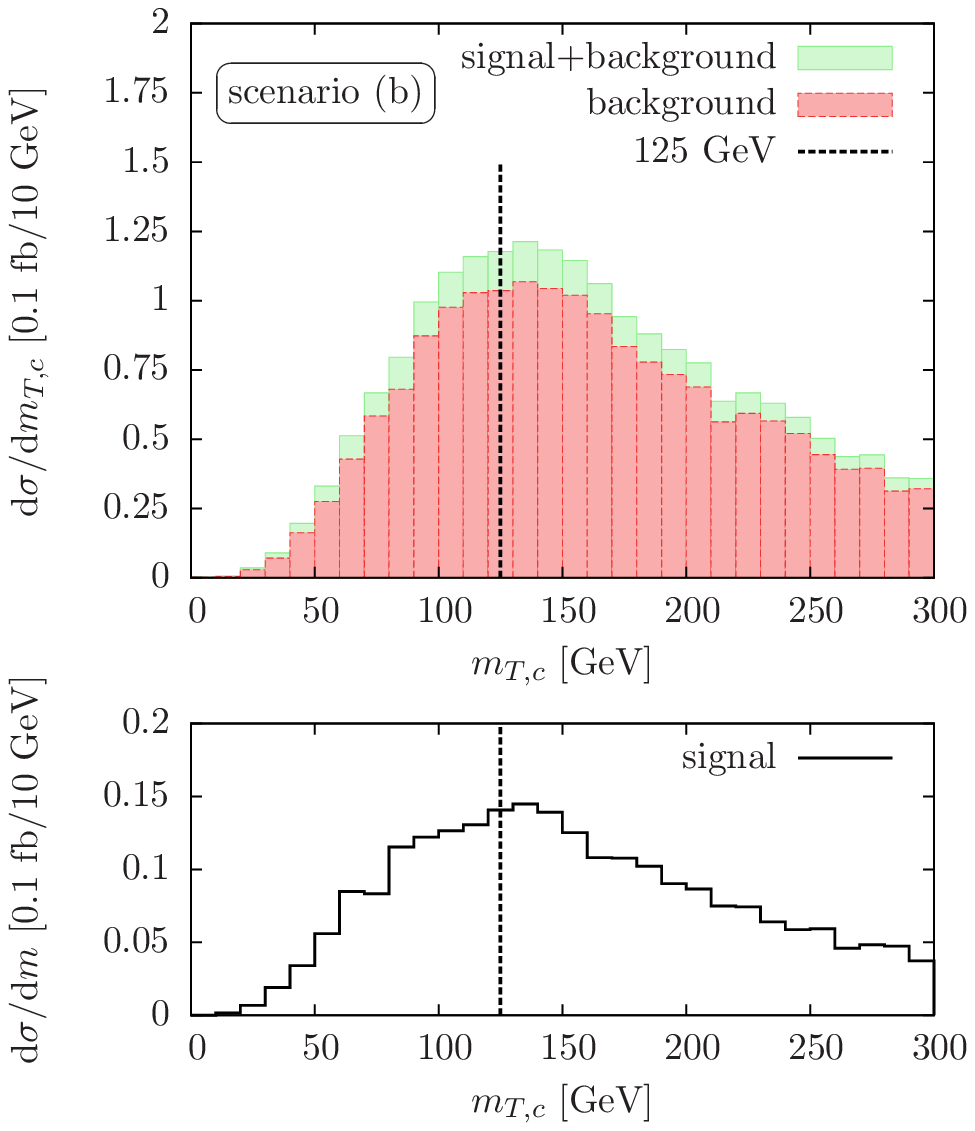}
    }\\
    \subfigure[][\label{fig:scen3}]{
       \includegraphics[width=0.4\textwidth]{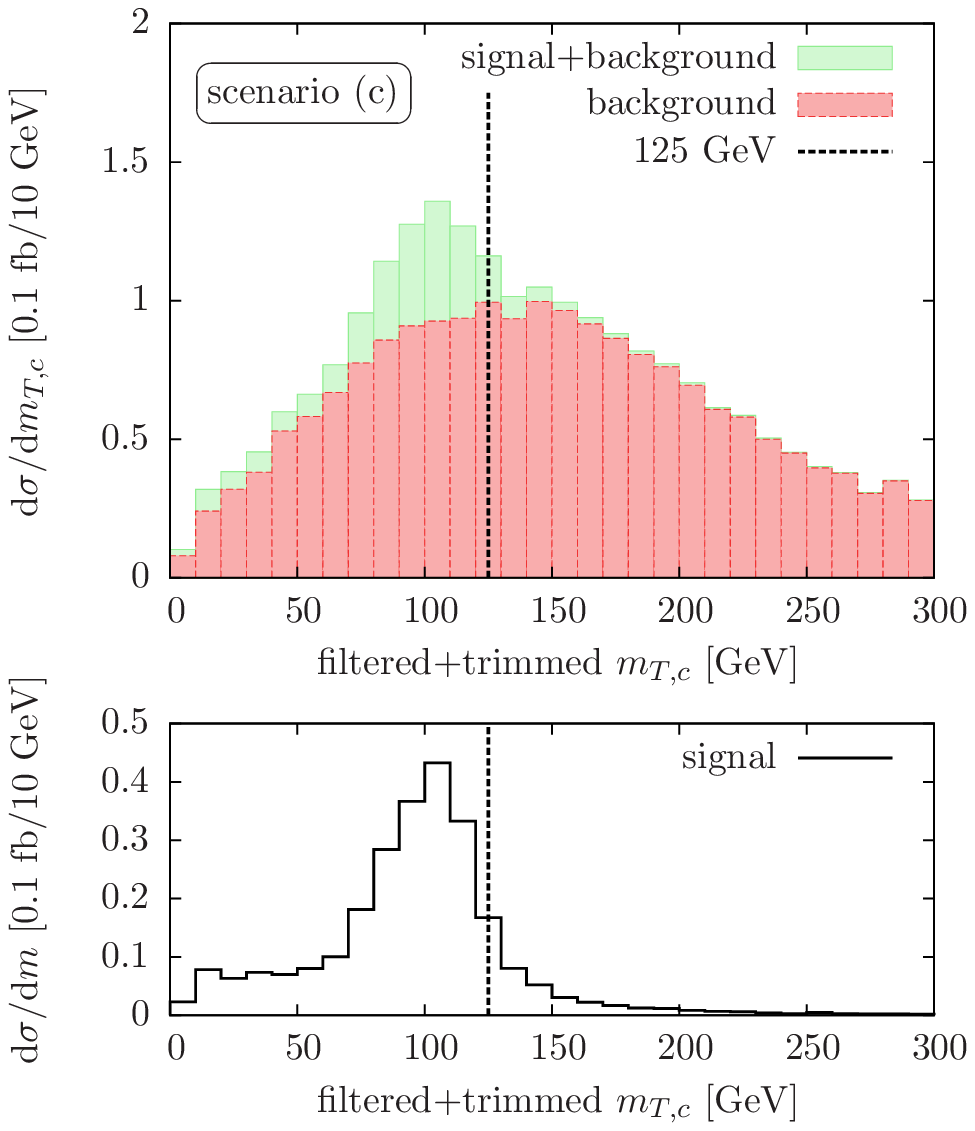}
     }
     \hspace{1.5cm}
    \subfigure[][\label{fig:scen4}]{
      \includegraphics[width=0.4\textwidth]{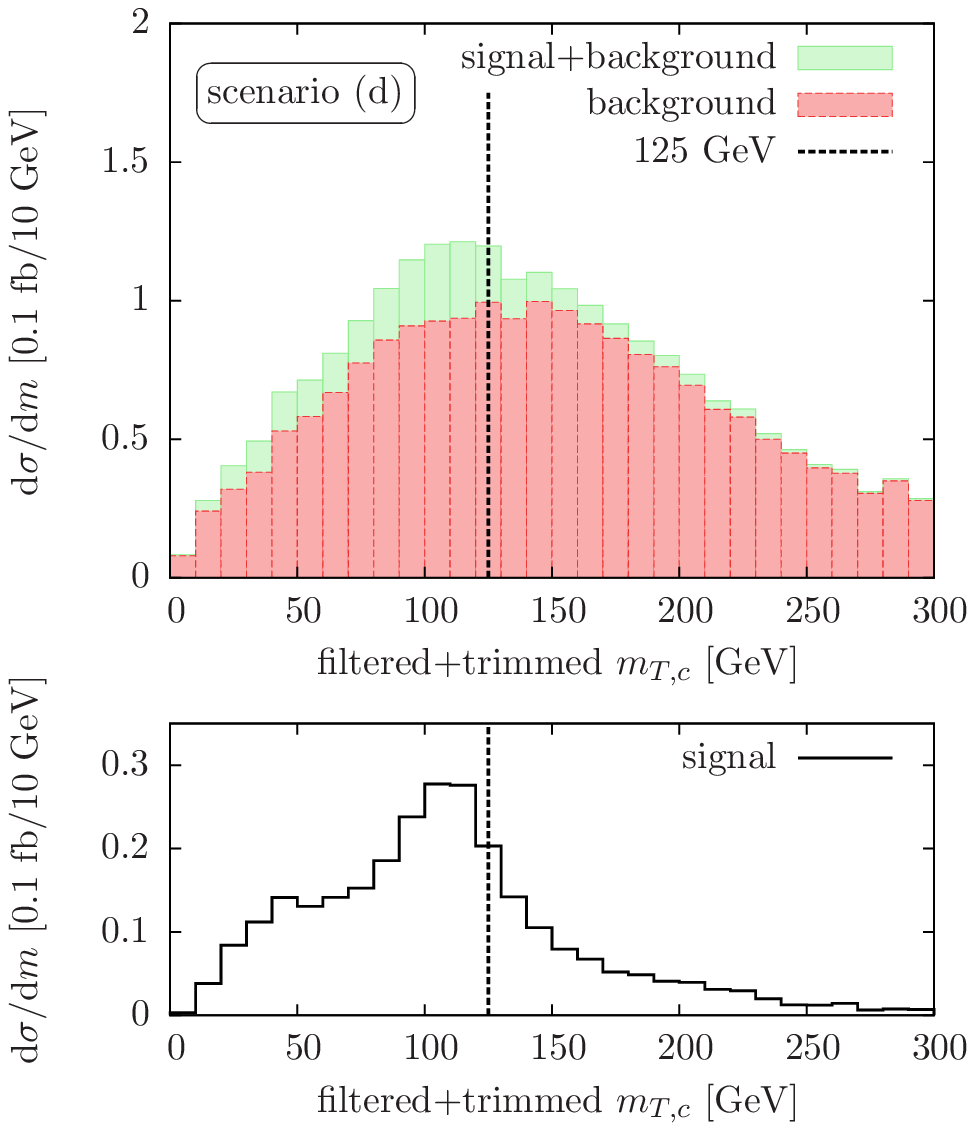}
    }\\
    \caption{\label{fig:result} Transverse mass or transverse cluster
      mass distributions after filtering and filtering+trimming,
      depending on the scenario of Fig.~\ref{fig:topref}. We assume
      $\text{BR}(\text{scenario~(i)})=1,~i=(a)-(d)$ for illustration purposes.}
\end{center}
\end{figure*}

Throughout the analysis we include detector smearing which we choose for jets
and leptons according to Ref.~\citeh{atlastdr}:
\begin{equation}
  \begin{split}
    \label{eq:resol}
    \text{jets}:\quad & {\Delta E\over E} = {5.2\over E} \oplus
    {0.16\over
      \sqrt{E}} \oplus 0.033\,,\\
    \text{leptons}:\quad & {\Delta {{E}} \over
      {E}} = {0.02}\,,\\
  \end{split}
\end{equation}
and we include the missing energy response from recent particle flow
fits of CMS~\citeh{pflow} (obtained from the 7 TeV data set,
Fig.~\ref{fig:reco}):
\begin{equation}
  \text{missing energy}:\quad  {\Delta {\slashed{E}}_T \over
    \slashed{E}_T} = {2.92 \over {\slashed{E}}_T} - 0.07\,.
\end{equation}
In fact, the quoted jet and lepton energy resolutions are better than
the ones obtained from particle flow observables~\citeh{pflow}. The
resolution of $\slashed{E}_T$ supersedes the original estimates of
Ref.~\citeh{atlastdr} and we can therefore assume that
Eq.~\gl{eq:resol} is a conservative baseline for the 14 TeV run at
luminosities at which the limits on the BRs will be stringent. We note
that the ${\slashed{E}}_T$ resolution is the main limiting factor of
the analysis that we pursue.

After applying all cuts, we are left with only minor possibilities to
enhance signal over background (apart from including an invariant mass
cut around the reconstructed Higgs mass, of
course). Fig.~\ref{fig:backgrounds} displays the $\slashed{E}_T$ and
fat jet $p_T$ distribution for the signal (assuming ${\text{BR}}=1$ to
the discussed topologies) and the discussed backgrounds. With all
described cuts applied, $t\bar t$ is sufficiently reduced, while the
electroweak $VV$+jet backgrounds are still large. Additionally, the
$\slashed{E}_T$ and $p_T$ distributions do not show any particular
discriminative features, which would allow to gain in $S/B$ from
tighter and harder missing energy and jet $p_T$ cuts (note that this
especially holds when we need to relax ${\text{BR}}=1$ for deriving
upper limits on the branching ratio). We find that SM-like Higgs
decays are sufficiently suppressed to make negligible contribution
to the background. 

\begin{figure*}[!t]
  \begin{center}
    \subfigure[][\label{fig:resultbra}]{
       \includegraphics[width=0.4\textwidth]{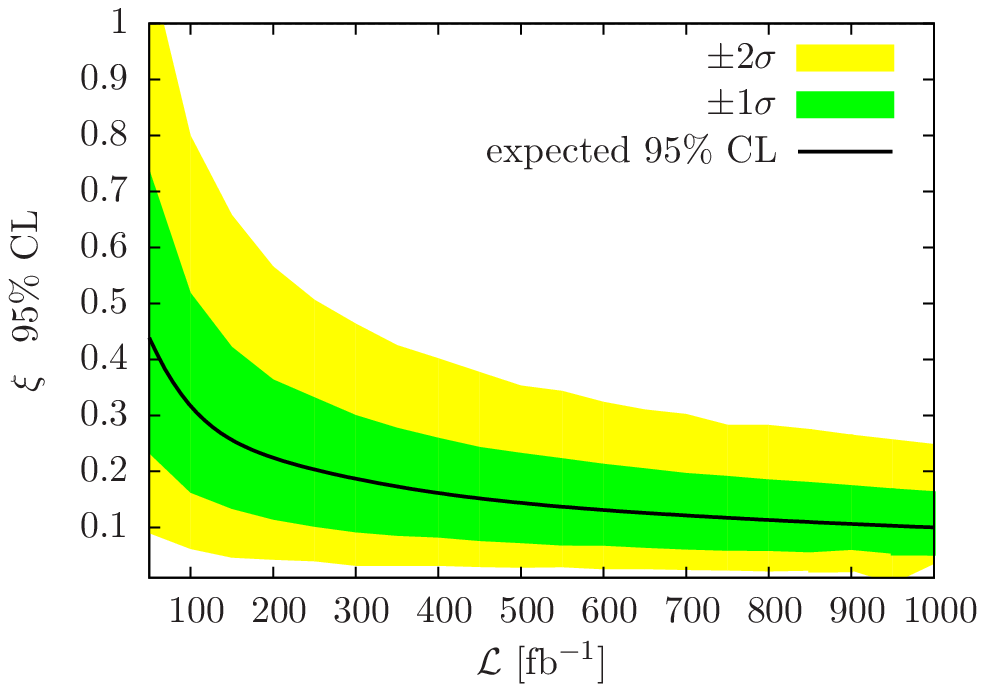}
     }
     \hspace{1.5cm}
    \subfigure[][\label{fig:resultbrb}]{
      \includegraphics[width=0.4\textwidth]{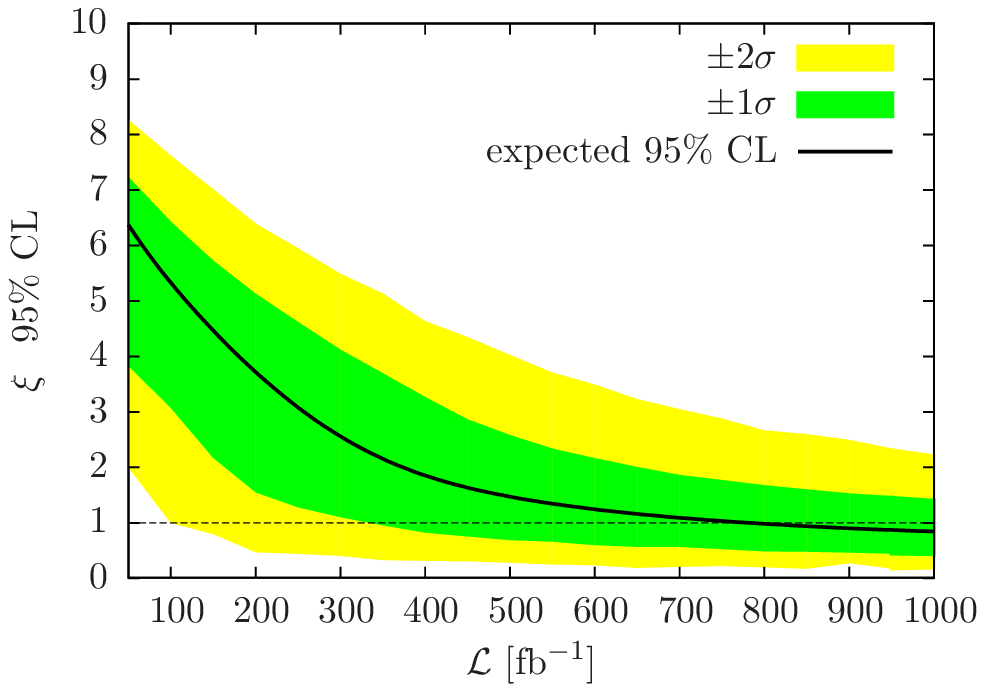}
    }\\
    \subfigure[][\label{fig:resultbrc}]{
      \includegraphics[width=0.4\textwidth]{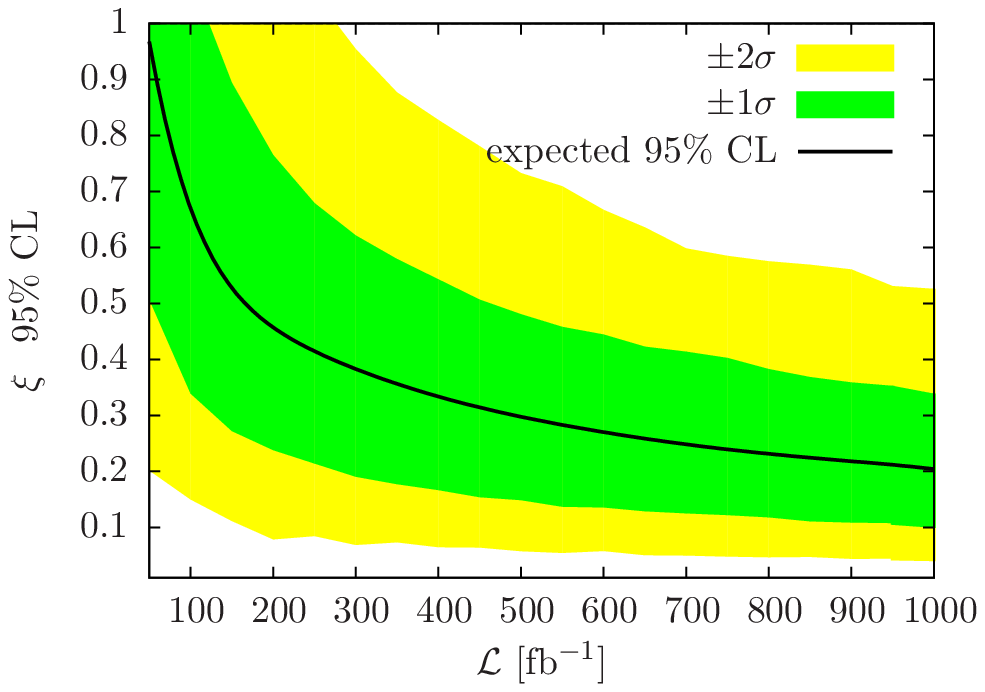}
    }
     \hspace{1.5cm}
    \subfigure[][\label{fig:resultbrd}]{
      \includegraphics[width=0.4\textwidth]{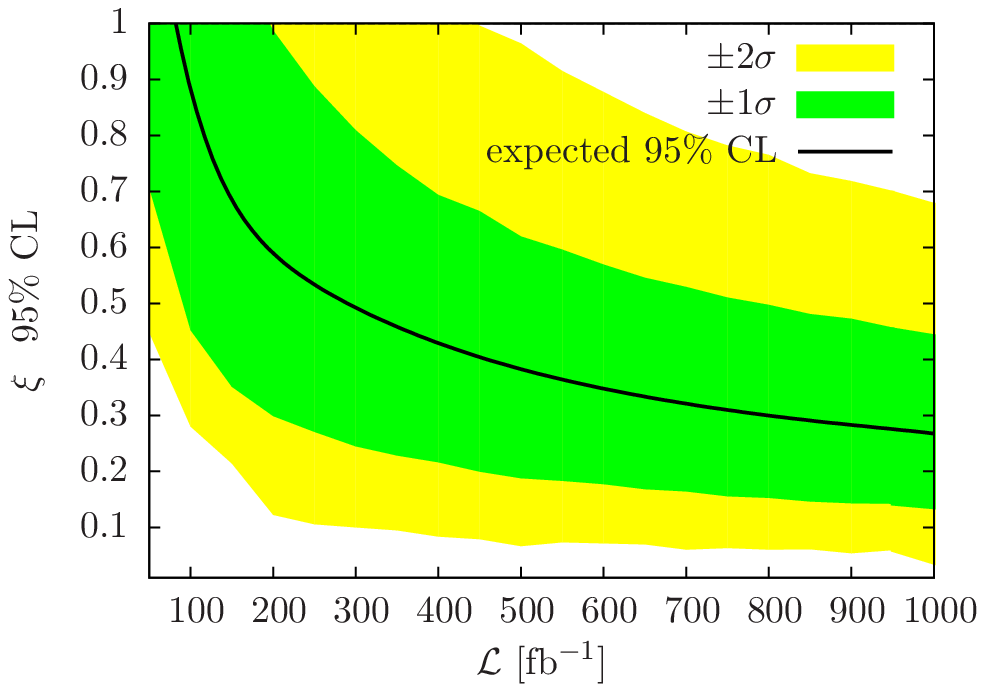}
    }\\
    \caption{\label{fig:resultbr} 95\% confidence level exclusion of
      the various scenario's branching ratio (or signal strength
      $\xi=\sigma\times \text{BR}/\sigma_{\text{SM}}$ to be more
      specific) from a direct measurement along the lines of
      Sec.~\ref{sec:analysis}. We assume a background template
      uncertainty of initially 10\%, which is we assume to saturated
      at 5\% at end-of-lifetime LHC luminosities of
      ${\cal{L}}=1000/{\text{fb}}$.}
\end{center}
\end{figure*}

In order to derive expected limits on the branching ratios we need to
reconstruct the $125~\gev$ resonance in this search channel in the
most efficient way. There are, in fact, dedicated observables, which
facilitate the reconstruction of invariant mass peaks when missing
energy is involved in the decay of a heavy resonance and they also
prove helpful in the present context. One example which
is useful for our purposes is the transverse mass definition
\begin{equation}
  m_T^2=\left(\sqrt{{\slashed{E}}_T^2+m_j^2} + \sqrt{{p_{T,j}^2 +
        m_j^2}}\right)^2 -   \left({\vec{p}}_{T,j} +
    {\slashed{\vec{p}}}_T \right)^2\,,
\end{equation}
which generalizes the invariant mass definition of Ref.~\citeh{dieter}
to the present situation. In Ref.~\citeh{dieter} an analogue
definition of the invariant mass is applied to $H\to\tau\tau$ (see
also~\citeh{batell}) with full leptonic $\tau$ decays in the collinear
approximation, {\it{i.e.}} the individual $\vec{\slashed{p}}$-lepton
correlation from each $\tau$ decay is incorporated. Applied to our
signatures this corresponds to a leptonic version of
Fig.~\ref{fig:topref}(a), and we can expect a good reconstruction of
the Higgs resonance for this decay scenario.

Another observable which reconstructs the mass of a heavy decaying
resonance from a kinematic endpoint measurement is the so-called
transverse cluster mass~\citeh{mtcluster}
\begin{multline}
  m^2_{T,c}= \left( \sqrt{ m^{2}_j+p_{T,j}^{2} }+
    \slashed{E}_{T} \right)^2 - \left( {\vec{p}}_{T,j}
    +\slashed{\vec{p}}_{T} \right)^2\,.
\end{multline}
This observable is a good choice if there is a relatively large amount
of visible energy compared to invisible
energy~\citeh{Englert:2011iz,Englert:2008tn}, as encountered in our
scenarios Fig.~\ref{fig:topref}(c) and (d), for which it allows to
reconstruct the Higgs mass from a Jacobian peak.

For signatures involving a lot of missing energy such as
Fig.~\ref{fig:topref}(c), we need to break the degeneracy of the
${\slashed{E}}_T$ vector to facilitate an approximate mass
reconstruction. For scenario Fig.~\ref{fig:topref}(b) we project
${\slashed{p}}$ onto the hardest reconstructed subjet after filtering
$\tilde j$, and assign an energy $\slashed{E}=E_{\tilde j}$. We define
a second missing energy four vector from the shift with respect to the
original missing energy vector
$\tilde{\slashed{p}}={\slashed{p}}_{\text{orig}}-\slashed{p}$. This
breaks the missing energy degeneracy and assumes a collinear
democratic $A$ decay in the lab frame. Resolution effects have the
biggest influence on this decay scenario and the mass reconstruction
will be worst among the four scenarios that we discuss in this work.

\bigskip 

The corresponding invariant mass distributions for our scenarios
Fig.~\ref{fig:topref}(a)-(d) after all cuts have been applied are
shown in Fig.~\ref{fig:result}, where we again assume
${\text{BR}}({\text{scenario~(i)}})=1,~i=(a)-(d)$. We pick the best
representation of the Higgs mass peak by studying the invariant mass
distributions before and after filtering and trimming. From the
different jet-substructure that follows as a consequence of the
different decay topologies, trimming can remove too much of the signal
when the light final state quark pairs are relatively widely separated
in scenario (a).

When focussing on scenarios with an increasingly large amount of
$\slashed{E}_T$ compared to hadronic activity, sensitivity
decreases. This is a consequence of the applied subjet analysis, which
requires by definition a significant amount of a hadronic
energy. Depending on the scenario we therefore become increasingly
sensitive to uncorrelated initial state radiation and underlying event
splash-in which degrades the mass resolution, while filtering and
trimming serve to suppress part of the pollution from underlying event.

\begin{figure*}[t]
  \begin{center}
    \parbox{0.4\textwidth}{
      \subfigure[][]{
      \includegraphics[width=0.4\textwidth]{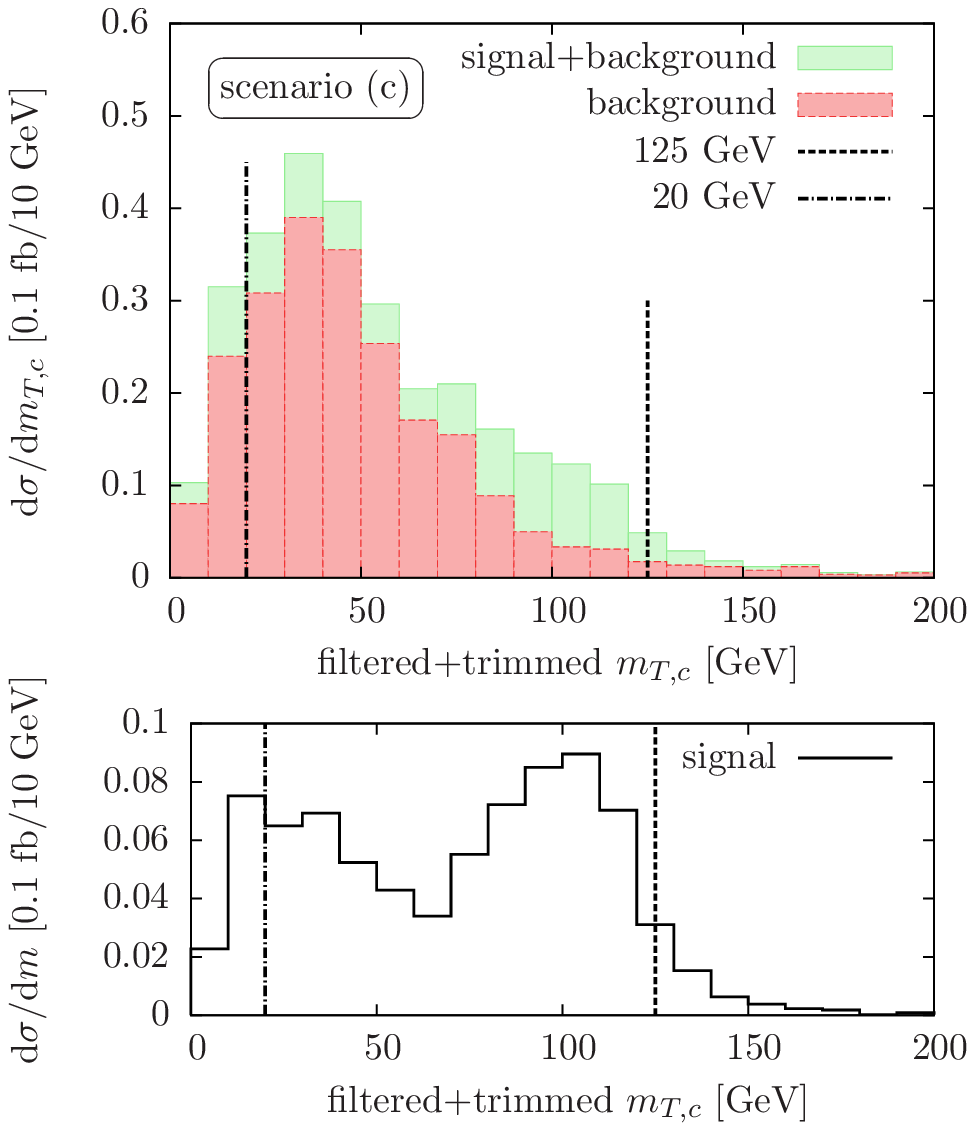}}
    }
    \hspace{1.5cm}
    \parbox{0.4\textwidth}{
      \subfigure[][]{
        \includegraphics[width=0.4\textwidth]{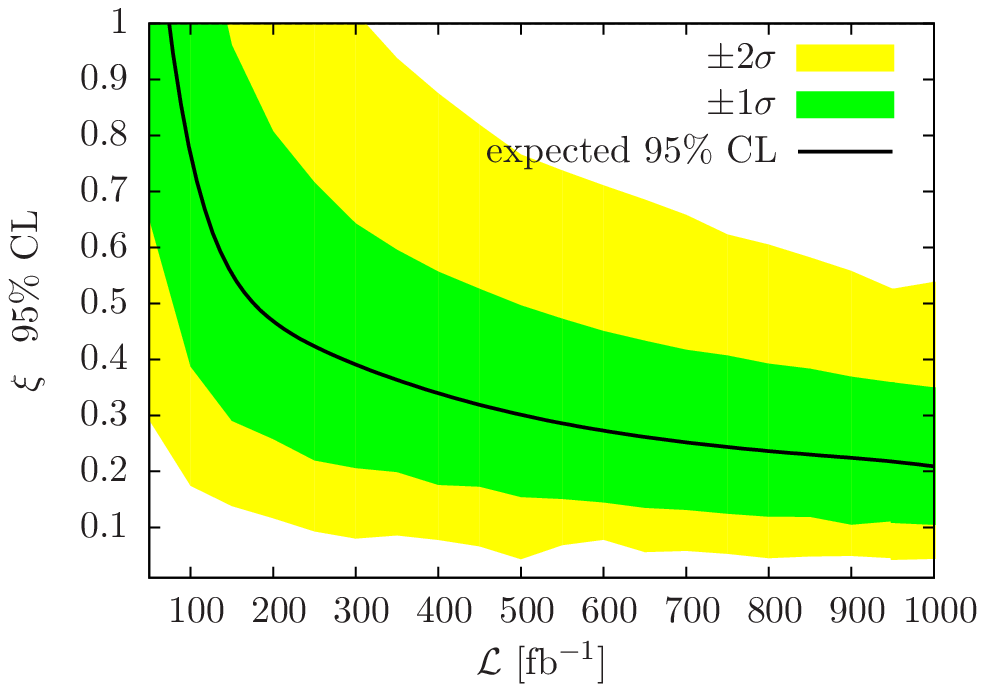}}
      }
      \caption{\label{fig:resultstrin} (a) Transverse cluster mass for
        scenario (c) with the additional requirement
        Eq.~\gl{eq:addcut}, and (b) the resulting branching ratio
        extraction along the lines of Sec.~\ref{sec:bounds}.}
  \end{center}
\end{figure*}

\section{Branching ratio limits at the LHC}
\label{sec:bounds}

We move forward and use the distributions of Fig.~\ref{fig:result} to
derive an estimate of the bounds on the branching ratios that can be
obtained at the LHC $14$~TeV from direct measurements. We follow the
experimentalists' approach and use the histograms of
Fig.~\ref{fig:result} as an input to the CL$_s$ method~\citeh{junk,
  Read:2002hq} to derive 95\% CL$_s$ exclusion limits on the branching
ratio using the binned log-likelihood test statistic. We include a
shape uncertainty of the background template in the computation of
confidence levels that eventually enter the CL$_s$ ratio. We choose this
shape uncertainty as a flat profile which we assume dependent on the
integrated luminosity 
\begin{equation}
  \Delta({\cal{L}})=10\%-5\% \sqrt{ {\cal{L}}
    /1000~{\text{fb}}^{-1}}
\end{equation}
to mimic an improved understanding of the
measurement and theoretical uncertainty in due time.

The expected 95\% confidence level exclusion is shown in
Fig.~\ref{fig:resultbr}. Direct measurements yield results
${\text{BR}}\lesssim 0.1,\dots,1$, and are highly sensitive to the
amount of hadronic energy that we observe in the final state and the
correlation of $\slashed{E}_T$ with the visible part of the final
state.

We learn from Fig.~\ref{fig:scen2} that the combination of little
hadronic energy, trigger criteria and detector effects together with
initial-state-radiation pollution results in an extremely challenging
signature at the hadronically busy LHC environment. As a direct
consequence, only weak limits can be obtained for decay scenario
(b). This is mostly due to the very limited possibilities to improve
$S/B$ and the poor mass resolution that can be obtained from a
signature that predominantly governed by missing energy in a channel
whose diboson backgrounds' also contain a sizable amount of
${\slashed{E}}_T$.
 
Turning to final states that contain more hadronic energy
Fig.~\ref{fig:topref}(c) and (d), we can impose stronger limits from
direct measurements (${\text{BR}}\lesssim 0.3$).  An optimal choice of
observables adopted to the specifics of Fig.~\ref{fig:topref}(a) in
comparison to (c) can push the bounds to the 10\% level.

\bigskip

An alternative route for putting limits on scenario (c), where
hadronic energy correlations in the fat jet's substructure can be
resolved, is exploiting the fat jet's ``active area'' (see also
Ref.~\citeh{Englert:2011iz}) in a way that also incorporates the
correlation with the missing energy. This is straightforwardly
achieved by imposing an additional cut on the ratio of the fat jet's
transverse momentum and the events reconstructed transverse cluster
mass $m_{T,c}({\slashed{E}},j)$
\begin{equation}
  \label{eq:addcut}
p_{T,j}/m_{T,c}>2\,.
\end{equation}
The resulting $m_{T,c}$ is shown in Fig.~\ref{fig:resultstrin}
together with the projected BR extraction. While the reduction in
statistics eventually compensates the enhancement in $S/B$ (compare
Fig.~\ref{fig:resultstrin}(b) to Fig.~\ref{fig:resultbr}(c)), we are
able to approximately reconstruct our light Higgs partner
$m_A=20~\gev$ from the jet substructure. This is only feasible by
exploiting visible final state energy correlations -- identical strategies
proof unsuitable for scenarios (a), (b) and (d).

A successful and intangible extraction of the light scalar mass from
the associated production channel hence depends on a good
understanding of the background at small $m_{T,c}$ and a sizable
branching ratio $H\to AA \to {\text{scenario~(c)}}$. Current fits
limit the invisible branching ratio to be ${\text{BR}}\lesssim
0.5$. Since all of our signatures are missed in standard searches, our
scenarios are constrained by this loose bound, so that there is the
possibility to find light scalars in that particular mass range in the
near future.

\section{Summary and Conclusions}
After the discovery of the Higgs-like resonance at around 125 GeV, a
further investigation of the resonance's compatibility with the SM
expectation is ongoing. A precise measurement of the resonance's
couplings to SM matter depends on the treatment of non-standard
partial decay widths, which is typically performed in global fit to
the quantitative \hbox{(non-)observation} in different exclusive Higgs final
states. Non-standard situations are hereby often treated as a
nuisance.

Hence, the necessity to investigate non-SM phenomenological situations
is imperative: If an indication of a non-zero branching ratio in a
specific channel is accessed in a direct measurement, the evidence for
physics beyond the SM will be accompanied by an insight in its
particular realization as opposed to a mere parametrization of
$\Gamma_{\text{inv}}$ in Eq.~\gl{decaywidth} as part of a global
fit. This inevitably comes at the price of a decreased sensitivity
({\it{i.e.}} less stringent limits) which results from investigating a
single channel as opposed to combinations and correlations of many.

\bigskip

We have investigated the possibility to constrain non-standard
branching ratios via direct measurements for a broad range of
effective models, which contain both invisible and visible final state
particles in this paper. Gaining sensitivity to these phenomenological
situations is characterized by a number of shortcomings, that range
from trigger issues all the way to the Higgs mass reconstruction at a
reasonable signal-over-background ratio. Since trigger thresholds are
typically bound to hardware specifications, we devise an analysis
strategy which builds upon associated production and a boosted Higgs
final state. As a consequence, we need to deal with an event topology
where missing energy is collimated with the visible Higgs decay
products in decays $H\to AA$ with $2 m_A \ll 125~\gev$ (a choice
$2m_A \sim 125~\gev$ exhibits too much $\slashed{E}_T$-jet
decorrelation to facilitate signal reconstruction on top of a large
background rejection). Recent investigations targeting the
reconstruction of $\slashed{E}_T$ based on particle-flow as well as
the extraction of fake $\slashed{E}_T$ from a combination of theory
and experiment suggest that systematics can be brought under
sufficient control to perform such analyses in this particular
channel.

We find that, depending on the specific realization of the partially
visible Higgs decay (which is characterized by different
${\slashed{E}}_T$-jet substructure correlations) bounds as good as
${\text{BR}}\lesssim 10\%$ can be obtained from a measurement of
boosted Higgs final states, applying a rather generic cut set up which
serves to obtain a reasonable $S/B$. Constraints can subsequently be
obtained from an optimal choice of observable which, depending on the
amount of visible final state energy, also allows to partially
reconstruct parts of the Higgs' potential decay chain.

\bigskip

\noindent {\it{Acknowledgements}} --- CE acknowledges funding by the Durham
International Junior Research Fellowship scheme. This work was supported in part
by the STFC.


\end{document}